\documentstyle[epsf,psfig]{elsart}
\begin{document}
\begin{frontmatter}
\title{Formation of correlations in strongly coupled plasmas}
\author[Rostock]{K. Morawetz,}
\author[Prag]{V\'aclav \v Spi\v cka and Pavel
Lipavsk\'y}
\address[Rostock]{Fachbereich Physik, Universit\"at Rostock,
18051 Rostock, Germany}
\address[Prag]{Institute of Physics, Academy of Sciences, Cukrovarnick\'a 10,
16200 Praha 6, Czech Republic}

\begin{abstract}
The formation of binary correlations in plasma is studied from the
quantum kinetic equation. It is shown that this formation is much
faster than dissipation due to collisions. In a hot (dense) plasma the
correlations are formed on the scale of inverse plasma frequency (Fermi
energy). We derive
analytical formulae for the time dependency of the potential energy which 
measures the extent of correlations. We discuss the 
dynamical formation of screening and compare with the statical screened 
result. Comparisons are made with molecular dynamic simulations. 
\end{abstract}
\end{frontmatter}

\begin{figure}
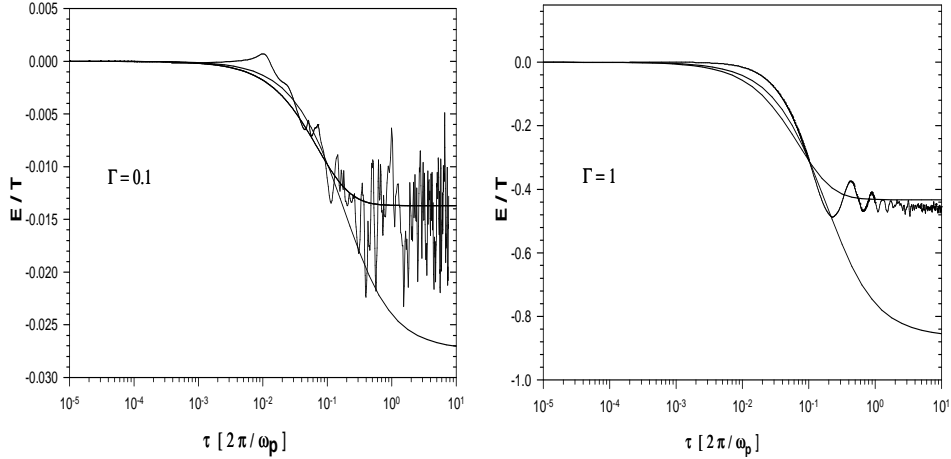

\parbox[t]{6cm}{
\psfig{figure=z01.epsi,width=6cm,height=6cm,angle=0}}
\hspace{1ex}
\parbox[t]{6cm}{
\psfig{figure=z1.epsi,width=6cm,height=6cm,angle=0}}
\caption{\label{1}The formation of correlation energy due to molecular
dynamic simulations \protect\cite{ZTRa95} together with the result of
(\protect\ref{v2}) for a plasma parameter $\Gamma=0.1$ (left) and
$ \Gamma=1$ (right). The upper curve is the static and
the lower the dynamical calculation of (\ref{v2}).
The latter one approaches the Debye-H\"uckel result.}
\end{figure}

Recent lasers allow to create a high density plasma within few
femto seconds and observe its time evolution on a comparable scale
\cite{HJ96,THWS96}. In this
paper we discuss the very first time regime, the transient
regime, in terms of the energy balance.
Let us assume a typically set up of molecular dynamics. One takes $N$
particles, distributes them randomly into a box and let them
classically move under Coulomb forces due to their own charges.
Their first movement thus forms correlations which lower
the Coulomb energy $V_{\rm C}=e^2/r$. This build up of screening
stops when the effective Debye potential
$V_{\rm D}=e^2 {\rm e}^{-\kappa r} /r$ is reached. We will
discuss the formation of correlations in terms of correlation
energy.
To this end we can use a kinetic equation, 
which leads to the total energy conservation. It is
immediately obvious that the ordinary Boltzmann equation cannot be
used because the kinetic energy is an invariant
of its collision integral. We have to consider
non-Markovian kinetic equations of Levinson type \cite{HJ96}
\begin{eqnarray}
\frac{\partial}{\partial t}f_a(t)&=&\frac{2}{\hbar^2}\sum\limits_b
\int\frac{dpdq}{(2\pi\hbar)^6}V_{\rm D}^2(q)
\int\limits_0^t d\bar t\,
\exp\left\{-{t-\bar t\over\tau}\right\}\,
{\rm cos}\left\{\frac{1}{\hbar}(t-\bar t)\Delta_E\right\}
\nonumber\\
&&\times\left\{\bar f'_a\bar f'_b(1\!-\!\bar f_a)(1\!-\!\bar f_b)-
\bar f_a\bar f_b(1\!-\!\bar f'_a)(1\!-\!\bar f'_b)\right\},
\label{kinetic}
\end{eqnarray}
where $\Delta_E={k^2\over 2m_a}+{p^2\over 2m_b}-{(k-q)^2\over 2m_a}-
{(p+q)^2\over 2m_b}$ denotes the energy difference between initial and final
states. The retardation of distributions, $\bar f_a(k,\bar t)$,
$\bar f'_a(k-q,\bar t)$ etc., is balanced by the lifetime $\tau$. The
total energy conservation for Levinson's equation has been
proved in \cite{M94}.
The solution in the short-time region $t\ll\tau$
can be written down analytically. In this time domain we can neglect the
time evolution of distributions, $\bar f_a(\bar t)=f_a(0)$, and the life-time
factor, $\exp\left\{-{t-\bar t\over\tau}\right\}=1$. The resulting expression for (\ref{kinetic}) describes then how two particles correlate their motion to avoid the
strong interaction regions.
This
very fast formation of the off-shell contribution to Wigner's
distribution has been found in numerical treatments of Green's
functions \cite{D841,K95}.
Of course, starting with a sudden switching approximation we have Coulomb
interaction and during the first transient time period the screening is
formed. This can be described by the non-Markovian Lenard - Balescu
equation \cite{Moa93} instead of the static screened equation (\ref{kinetic})
leading to the dynamical expression of the correlation energy [details, see \cite{MLSa97}].
To demonstrate its
results and limitations, we use Maxwell
initial distributions at the high temperature limit, where the distributions are
non-degenerated.
From (\ref{kinetic}) we find with ${\partial \over \partial t} E_{\rm corr}=-
\sum_a\int{dk\over(2\pi\hbar)^3}{k^2\over 2m_a}{\partial \over \partial t} f_a$
\begin{eqnarray}
{\partial \over \partial t} {E_{\rm corr}^{\rm static}(t) \over n}&=& -{e^2 \kappa T\over 2 \hbar}{\rm Im}
\left [(1+2 z^2 ) {\rm e}^{z^2} (1- {\rm erf} (z)) -{2 z \over \sqrt{\pi}} \right ] \nonumber\\
{\partial \over \partial t} {E_{\rm corr}^{\rm dynam}(t) \over n}&=&
-{e^2 \kappa T \over  \hbar}{\rm Im}
\left [{\rm e}^{z_1^2} (1- {\rm erf} (z_1)) \right ]
\label{v1}
\end{eqnarray}
where we used $z =\omega_p \sqrt{t^2 - i t {\hbar \over T}}$ and $z_1 =\omega_p \sqrt{2 t^2 - i t {\hbar \over T}}$.
This is the analytical quantum result of the time derivative of the formation of correlation for statically as well as
dynamically screened potentials. For the classical limit we are
able to integrate expression (\ref{v1}) with respect to times and arrive at
\begin{eqnarray}
E_{\rm corr}^{\rm static}(t)&=&-{1\over 4}e^2n\kappa
\Biggl\{1+{2\omega_p t\over\sqrt{\pi}}
 -\left(1+2\omega_p^2t^2\right)\exp\left(\omega_p^2t^2\right)
\left[1-{\rm erf}(\omega_p t)\right]\Biggr\}\nonumber\\
E_{\rm corr}^{\rm dynam}(t)&=&-{1\over 2}e^2n\kappa
\Biggl\{1-\exp\left({\omega_p^2 \over 2 } t^2\right)
\left[1-{\rm erf}({\omega_p \over \sqrt{2}} t)\right]\Biggr\}.
\label{v2}
\end{eqnarray}
In Figs.~\ref{1}, this formulae are compared with molecular
dynamic simulations \cite{ZTRa95} for two values of the plasma parameter
$\Gamma=0.1$ and 1. This parameter $\Gamma={e^2\over a_eT}$, where
$a_e=({3\over 4\pi n})^{1/3}$ is the inter-particle distance or
Wigner-Seitz radius, measures the strength of the Coulomb coupling.
Ideal plasma are found for $\Gamma\ll1$. In this region the static formula
(\ref{v2}) well follows the major trend of the numerical result, see
Fig.~\ref{1}. The agreement is in fact surprising, because the static result
underestimates the dynamical long time result of
Debye- H\"uckel $\sqrt{3} /2 \Gamma^{3/2}$ by a factor of two, which can be seen from the long time and classical limit
$b^2=(\hbar\kappa)^2{m_a+m_b\over 8m_am_b T}\to 0$
\begin{eqnarray}
E_{\rm corr}^{\rm dynam}(\infty)=-{e^2 \kappa \over 2
}{\sqrt{\pi} \over b} (1-{\rm e}^{b^2} {\rm erfc}
(b))&=&-{1\over 2}e^2n\kappa+o(b)
\nonumber\\
E_{\rm corr}^{\rm static}(\infty)=-{e^2 \kappa \over 4
} (1-\sqrt{\pi} \ {\rm erfc} (b))&=&-{1\over 4}e^2n\kappa+o(b).
\label{class}
\end{eqnarray}
The first result represents the Montroll correlation energy
\cite{kker86,RSWK95}.
The explanation for this fact is that we can prepare the initial configuration within our kinetic theory such that sudden switching of interaction is fulfilled. However, in the simulation experiment we have initial correlations which are due to the set up within quasiperiodic boundary condition and Ewald summations. This obviously results into an effective statically screened Debye potential, or at least the simulation results allow for this interpretation.

For $\Gamma=1$, see Fig.~\ref{1}, non-ideal effects become important and
the formation time is underestimated within (\ref{v2}).
This is due to non-ideality which was found to be an
expression of memory effects \cite{MWR93} and leads to a later
relaxation.

The characteristic time of formation of correlations at high temperature
limit is given by the inverse plasma frequency
$\tau_c\approx{1\over\omega_p}={\sqrt{2}\over v_{\rm th}\kappa}$.
The inverse plasma frequency indicates that the long
range fluctuations play the dominant role. 
This is equivalent to the time a particle needs to travel through the range 
of the potential with a thermal velocity $v_{\rm th}$. This confirms the
numerical finding of \cite{BKSBKK96} that the correlation or memory
time is proportional to the range of interaction.
In the low temperature region, i.e., in a highly degenerated system
$\mu\gg T$, one finds a different picture \cite{MK97,MSL97a}.
Unlike in the classical case, the equilibrium limit of the degenerated
case is rapidly built up and then oscillates around
the equilibrium value. We can define the build up time
$\tau_c$ as the time where the correlation energy reaches its first
maximum,
$\tau_c=1.0{\hbar\over\mu}$ with the Fermi energy $\mu$.
Note that $\tau_c$ is in agreement with the quasiparticle formation
time known as Landau's criterion.
Indeed, the
quasiparticle formation and the build up of correlations are two
alternative views of the same phenomenon.
The formation of binary correlations is very fast on the time scale of
dissipative processes. Under extremely fast external perturbations,
like the massive femto second laser pulses, the dynamics of binary
correlations will hopefully become experimentally accessible. 

We are grateful to G. Zwicknagel who was so kind as to provide the data
of simulations. Stimulating discussion with G. R{\"o}pke is
acknowledged. This project was supported by the BMBF
(Germany) under contract Nr. 06R0884, the Max-Planck Society,
Grant Agency of Czech
Republic under contracts Nos. 202960098 and 202960021,  and the EC Human Capital and
Mobility Programme.

\end{document}